\documentstyle[12pt]{article}

\newcommand{\be}{\begin{equation}}
\newcommand{\ee}{\end{equation}}

\newcommand{\vesig}{\mbox{\boldmath${\rm \sigma}$}}
\newcommand{\vetau}{\mbox{\boldmath${\rm \tau}$}}
\newcommand{\venabla}{\mbox{\boldmath${\rm \nabla}$}}

\def\fun#1#2{\lower3.6pt\vbox{\baselineskip0pt\lineskip.9pt
\ialign{$\mathsurround=0pt#1\hfil##\hfil$\crcr#2\crcr\sim\crcr}}}

\title{\large \bf  Chirality violating condensates in QCD   and their connection\\
with zero mode solutions of quark Dirac equations  }
\author{B.L.Ioffe\\
Alikhanov Institute of Theoretical and Experimental Physics,\\ B.Cheremushkinskaya 25,
117218, Moscow, Russia}

\begin{document}
\date{}

\maketitle

\begin{abstract}

 It is demonstrated, that chirality violating  condensates in massless QCD arise entirely
 from
 zero mode solutions of Dirac equations in arbitrary gluon fields. The model is
 suggested, where the zero mode solutions are the ones for quarks, moving in the
 instanton field. Basing on this model were calculated the quark condensate magnetic
 susceptibilities of dimensions  $3(\chi)$ and 5 ($\kappa$ and $\xi$). The good considence
 of the values $\chi,\kappa$ and $\xi$, obtained  in this approach  with ones, found
 from the hadronic spectrum ia a serious argument in favour, that instantons are the
 only source of chirality violating condensates in QCD.
The temperature dependence of the quark condensate is discussed. It is shown that the
phase transition, corresponding to the $T$-dependence of the quark condensate $\alpha(T)$
as an order parameter, is of the type of crossover.

\bigskip

Keywords: zero mode, instanton, condensat

\bigskip

12.38. Aw, 11.30. Rd, 11.15. Tk

\end{abstract}

\vspace{1cm}

It is well known, that because of small values of light quark masses the perturbative QCD
pocesses the property of chiral symmetry in cases, when the heavy quarks contributions
can be neglected. However, in the real hadronic world the chiral symmetry is badly
violated. This statement evidently follows from  the existence of large proton mass and
the absence of the negative parity baryon, degenerate with proton. It is also well known
that large values of light quark condensates indicate the violation of chiral symmetry in
QCD. These  two facts are deeply interconnected: the values of the proton mass can be
expressed through  the value of quark condensate \cite{Ioffe}.

In this talk I discuss the appearance of quark condensate in nonperturbative QCD. It is
established the connection of quark condensate value with zero mode solutions of quark
Dirac equation \cite{IoffeBL}. In this way it is clarified the nature of proton mass and,
as a consequence,  the nature of all visible mass in the Universe \cite{UFN}.

 Consider QCD action in Euclidean space-time
\be S =\frac{1}{4} \int d^4 x G^2_{\mu\nu} -\int d^4 x \sum\limits_f \biggl [\psi^+_f
(i\gamma_{\mu} \nabla_{\mu} + im_f)\psi_f\biggr]\label{1a}\ee where $G^n_{\mu\nu}$ is
gluon field tensor, the sum is over quark flavours.

\be \nabla_{\mu} =\partial_{\mu} +ig \frac{\lambda^n}{2} A^n_{\mu}\label{2a}\ee and
$A^n_{\mu}$ is the gluon field. Pay attention, that in Euclidean formulation of QCD
$\overline{\psi}$ is replaced by $\psi^+$. (The review of Euclidean formulation of QCD
and instantons is given in \cite{Inst},  see especially \cite{Vain}.) The Dirac equation
for massless quark in Euclidean space time has the form:
\be -i\gamma_{\mu} \nabla_{\mu} \psi_n(x) =\lambda_n\psi_n(x)\label{3a}\ee where
$\psi_n(x)$ and $\lambda_n$ are the eigenfunctions and eigenvalues of the Dirac operator
$-\venabla=-i\gamma_{\mu} \nabla_{\mu}$. Expand the quark fields operators into the left
and right ones
$$ \psi =\frac{1}{2} (1+\gamma_5) \psi_L +\frac{1}{2} (1-\gamma_5)\psi_R$$
\be \psi^+ = \psi^+_L\frac{1}{2} (1+\gamma_5) +\psi^+_R \frac{1}{2}
(1-\gamma_5),\label{4a}\ee where
\be \gamma_5\psi_L =\psi_L,~~~\gamma_5\psi_R =-\psi_R\label{5a}\ee Then for nonzero
$\lambda_n$ the Lagrangian and the action reduces to the sum of two terms
\be L = -\int\biggl [~\psi^+_L \venabla \psi_R +\psi^+_R \venabla \psi_L~\biggr ] d^4 x
\label{6a}\ee completely symmetric under interchange $L\longleftrightarrow R$. Therefore
the solutions of the equations for left and right quark fields are also the same -- the
states, constracted from left  and right quarks are  completely  symmetrical. This
conclusion was obtained for fixed gluon field. It is evident, that the averaging over the
gluon fields does not change  it. Quite different situation arises in case $\lambda_0
=0$. The contribution of this term to  the Lagrangian:
\be \Delta L =\int d^4 x  [~\psi^+_L +\psi^+_R ~ ]\venabla \psi_0\label{7a}\ee is equal
to zero and no conclusion can be done about the symmetry of states build from left and
right quark fields. One of the consequences  from the said above  is that all chirality
violating vacuum condensates in QCD arise from zero mode solutions of Dirac equations
(\ref{3a}).

These general arguments are supported by the well known facts:\\
1. The general representation of the trace of quark  propagator  $S(x)$ is  expressed
through the spectral function $\rho(\lambda)$ as a function of eigenvalues $\lambda$
(K$\ddot{\mbox{a}}$llen-Lehmann representation):
\be Tr S(x^2) =\frac{1}{\pi} \int d\lambda \rho(\lambda)\Delta(x^2,\lambda)\label{3}\ee
At $x^2=0$ $\Delta(x^2,\lambda)$ reduces to $\delta(\lambda)$ and we have (in Minkowski
space-time):
\be \rho(0) =-\pi \langle 0\mid \overline{\psi}(0)\psi (0)\mid 0\rangle.\label{4}\ee (The
Banks-Casher relation \cite{Banks}).\\
2. The zero-mode solution of (\ref{3a}) for massless quark in the instanton field is the
right  wave function -- $\psi_R(x) = (1-\gamma_5)\psi(x)$ and in the field of
anti-instanton is the left one -- $\psi_L(x) = (1+\gamma_5)\psi(x)$ \cite{Brown,Jackiw}.

Basing on the statements, presented above, let us formulate the model for calculation of
chirality violating  vacuum condensates in QCD. Suppose, that vacuum expectation   value
(v.e.v.) of the chirality violating operator $O_{c.v.}$ is proportional to matrix element
$\psi^+_0 O_{c.v.}\psi_0$, where $\psi_0$ is the zero-mode solution of Eq.(\ref{3a}) in
Euclidean space-time:
\be \langle 0\mid \overline{\psi}O_{c.v.} \psi \mid 0\rangle \sim \psi^+_0
O_{c.v.}\psi_0.\label{5}\ee $\psi_0$ depends on $x$, on the position of the center of the
solution $x_c$, as well as on its size $\rho$: $\psi_0=\psi_0(x-x_c,\rho$). Eq.(\ref{5})
must be integrated over $x_c$,  what is equivalent to integration over $x-x_c$. (In what
follows the notation $x$ will be used for $x-x_c$.) We assume, that $\rho=$Const and find
its value from comparison with the known v.e.v.'s. Finally, introduce in (\ref{5}) the
coefficient of proportionality $n$. So, our assumption has the form:
\be \langle 0\mid \overline{\psi}(0) O_{c.v.} \psi(0)\mid 0\rangle =-n \int d^4 x
\psi^+_0(x,\rho)O_{c.v.} \psi_0(x,\rho)\label{6}\ee Our model is similar to delute
instanton gas model \cite{Schafer}, where $x_c$ is the position of instanton center.
Unlike the latter, where the instanton density has dimension 4, $n$ has dimension 3 and
may be interpreted as the density of zero-modes centers in 3-dimension space. Note, that
the left-hand side of (\ref{6}) is written in the Minkowski space-time, while the
right-hand side in Euclidean ones. (The sign minus is put in order to have $n$ positive.)
In perturbative calculation in case, when the number of flavours is more then 1, $N_f
>1$, the contribution of instantons to the action is suppressed by the factor, proportional to the product of $N_f -1$
light quark masses. Therefore, the assumption (11) means that such suppression is absent
in nonperturbative calculation. For $x$ and $\rho$-dependens of $\psi_0(x,\rho)$ we take
the form of the zero-mode solution in the field of instanton in $SU(2)$ colour group:
\be \psi_0(x,\rho) =\frac{1}{2}(1-\gamma_5) \frac{1}{\pi}
\frac{\rho}{(x^2+\rho^2)^{3/2}}\chi_0,\label{7}\ee where $\chi_0$ is the spin-colour
isospin $(\mid {\bf T} \mid =1/2$) wave function, corresponding to the total spin ${\bf
I+T=J}$ equal to zero, $J=0$. $\psi_0(x,\rho)$  is normalized to 1:
\be \int d^4 x \psi^+(x,\rho)\psi(x,\rho)=1\label{8}\ee

Consider first the quark condensate $\langle 0\mid \bar{q}q\mid 0\rangle$, the most
important chirality violating v.e.v., determining the values of baryon masses
\cite{Ioffe}-\cite{BLIoffe}. (Here $q=u,d$ are the fields of $u,d$-quarks). In this case
$O_{c.v.}=1$ and in accord with (\ref{6}),(\ref{8}) we have
\be n=-\langle 0\mid \bar{q}q\mid 0\rangle = (1.65 \pm 0.15) \times
10^{-2}~\mbox{GeV}^3~(\mbox{at 1 GeV})~[10]\label{9}\ee (The integration  over $SU(2)$
subgroup position in $SU(3)$ colour group as well as anti-instanton contribution are
included in the definition of $n$.) The anomous dimension of quark condensate is equal to
4/9. According to (\ref{9}) $n$ has the same anomalous dimension.  The size $\rho$ of the
zero-mode wave function can be found by calculation in the framework of our model of the
v.e.v.
\be -g \langle 0 \mid \overline{\psi} \sigma_{\mu\nu} \frac{\lambda^n}{2} G^n_{\mu\nu}
\psi \mid 0\rangle \equiv m^2_0 \langle 0\mid \bar{q} q \mid 0\rangle.\label{10}\ee The
parameter $m^2_0$ is equal  to \cite{Belyaev}: $m^2_0 =0.8$ GeV$^2$ at 1 GeV. The $m^2_0$
anomalous dimension is equal to -14/27. Working in the $SU(2)$ colour group, substitute
$\lambda^n$ by $\tau^a(a=1,2,3)$ and take for $G^a_{\mu\nu}$ the instanton field
\be G^a_{\mu\nu} (x,\rho) =\frac{4}{g} \eta_{a\mu\nu} \frac{\rho^2}{(x^2+\rho^2)^2},
\label{11}\ee where the parameter $\eta_{a\mu\nu}$ were defined by $'$t Hooft
\cite{Hooft} (see also \cite{Vain}). The substitution of (\ref{7}) and (\ref{11}) into
(\ref{6}) gives after simple algebra
\be \frac{1}{2} n \frac{1}{\rho^2} = m^2_0 n.\label{12}\ee Therefore, \be \rho
=\frac{1}{\sqrt{2} m_0} = 0.79 ~\mbox{GeV}^{-1} =0.256 \mbox{fm} (\mbox{at~ 1
GeV}).\label{13}\ee This value of $\rho$ is close to ones, used in the delute instanton
gas or instanton liquid models.

We are now in a position to  calculate less well known quantities -- the magnetic
susceptibilities of quark condensate, induced  by external constant  electromagnetic
field. \\The dimension 3 quark condensate magnetic susceptibility is defined by
\cite{Smilga}:
\be \langle 0 \mid \bar{q} \sigma_{\mu\nu} q \mid 0 \rangle_F =e_q \chi \langle 0\mid
\bar{q}q \mid 0 \rangle F_{\mu\nu},~~q = u,d,\label{14}\ee where quarks are considered as
moving in external constant weak electromagnetic field $F_{\mu\nu}$ and $e_q$ is the
charge of quark $q$ in units of proton charge (the proton charge $e$ is included in the
definition of $F_{\mu\nu}$). The left-hand side of (\ref{14})  violates chirality, so it
is convenient to separate explicitly  the factor $\langle 0 \mid \bar{q} q \mid 0\rangle$
in the right-hand side. It was demonstrated in \cite{Smilga} that $\langle 0\mid
\bar{q}\sigma_{\mu\nu} q \mid 0\rangle_F$ is proportional to the charge $e_q$ of the
quark $q$. A universal constant $\chi$ is called the quark condensate magnetic
susceptibility.

Let us determine the value of $\chi$ in our approach. For this goal it is necessary  to
consider Eq.\ref{3a} in the presence of external constant electromagnetic field
$F_{\mu\nu}$ and to find the first order in $F_{\mu\nu}$ correction to zero mode solution
(\ref{7}). This can be easily done by representing $\psi$ as
\be \psi(x,\rho) =\psi_0(x,\rho) +\psi_1(x,\rho),\label{15}\ee where $\psi_0$ is given by
(\ref{7}) and $\psi_1$ represents the proportional to $F_{\mu\nu}$ correction. Substitute
(\ref{15}) in Eq.\ref{3a} added by the term of interaction with electromagnetic field,
neglect $\psi_1$ in this term and solve the remaining equation for $\psi_1(x,\rho)$). The
result is:
\be \psi_1(x,\rho)= \frac{1}{16} e_q \eta_{a\mu\nu} \sigma_a F_{\mu\nu} x^2 \biggl
(1+\frac{1}{2} \frac{x^2}{\rho^2}\biggr ) \psi_0 (x,\rho), \label{16}\ee where $\sigma_a$
are Pauli matrices. The matrix element $\psi^+ \sigma_{\mu\nu} \psi$ appears to be equal:
\be \psi^+ \sigma_{\mu\nu} \psi = -\frac{1}{2} e_q F_{\mu\nu} \psi^+_0 x^2 \biggl (
1+\frac{1}{2}\frac{x^2}{\rho^2}\biggl )\psi_0.\label{17}\ee (The properties of
$\eta_{a\mu\nu}$ symbols \cite{Hooft},\cite{Vain} were exploited.) The v.e.v. (\ref{14})
in the Minkowski space-time is given by:
\be \langle 0\mid \overline{\psi}\sigma_{\mu\nu} \psi\mid 0\rangle_F = e_q F_{\mu\nu}
n\frac{1}{\pi^2} \int d^4 x x^2 \biggl (1+\frac{1}{2} \frac{x^2}{\rho^2}\biggr )
\frac{\rho^2}{(x^2+\rho^2)^3}\label{18}\ee (The normalization condition (\ref{8}) for
$\psi_0(x,\rho)$ was used.) It is convenient to express $n$ through quark condensate by
(\ref{9}), use the notation $x^2=r^2$, where $r$ is the radius-vector in 4-dimensional
space. Then according to (\ref{14}) we have:
\be \chi =-\rho^2 \int\limits^{R^2}_0 dr^2 r^4 \biggl ( 1+ \frac{1}{2}
\frac{r^2}{\rho^2}\biggr ) \frac{1}{(r^2 +\rho^2)^3}\label{19}\ee The integral (\ref{19})
is quadratically divergent at large $r$. So, the cut-off $R$ is introduced. Its value can
be estimated in following way. The volume occupied by one zero-mode in  3-dimensional
space is approximately equal to $1/n$ (the volume of the Wigner-Seitz cell). So, for
cut-off radius square $R^2$ in four-dimensions we put
\be R^2 =\frac{4}{3} \biggl ( \frac{3}{4\pi n}\biggr )^{2/3} =7.92
\mbox{GeV}^{-2}\label{20}\ee where the factor $4/3$ corresponds to transition from 3 to 4
dimensions. The calculation of the integral (\ref{19}) at the values of parameters $\rho$
(\ref{13}) and $R^2$  (\ref{20}) gives
\be \chi = -3.52 \mbox{GeV}^{-2}\label{21}\ee The quark condensate magnetic
susceptibility was previously calculated by QCD sum rule  method
\cite{Kogan}-\cite{Rohrwild} and expressed through the masses and coupling constants of
mesonic resonances. The recent results are:
\be \chi (1 \mbox{GeV)} =-3.15 \pm 0.3 \mbox{GeV}^{-2}~[18];~~\chi (1~\mbox{GeV)} = -2.85
\pm 0.5 \mbox{GeV}^{-2}~[19]\label{22}\ee (The earlier results, obtained by the same
method, were: $\chi(0.5$ GeV) =$ - $5.7 GeV$^{-2}$ \cite{Kogan} and $\chi(1$ GeV) =\\
--4.4$\pm 0.4$ GeV$^{-2}$ \cite{Balit}.) The anomalous dimension of $\chi$ is equal to
-16/27. It was accounted in \cite{Kogan}-\cite{Rohrwild}, but not in the presented above
calculation. (In some of these papers, the $\alpha_s$-corrections and continuum
contribution, were also accounted.) One can believe, that the value (\ref{21}) refer to 1
GeV, because the value of quark condensate (\ref{9}) refer to this scale and also because
the scale 1 GeV is a typical scale, where, on the one hand, the zero-modes and quark
condensates are quite important (see, e.g. \cite{Iof}) and, on the other, the instanton
gas model is valid \cite{Schafer}. Since the integral is quadratically divergent it is
hard to estimate the accuracy of (\ref{21}). I guess, that it is not worse, than 30-50\%.
In the limit of this error the result (\ref{21}) is in an agreement with those found in
phenomenological approaches.

Turn now to quark condensate magnetic susceptibilities of dimension 5, $\kappa$ and $\xi$
defined in Ref.\cite{Smilga}
\be g\langle 0\mid \bar{q} \frac{1}{2}  \lambda^n G^n_{\mu\nu} \bar{q}\mid 0 \rangle_F =
e_q \kappa F_{\mu\nu} \langle 0 \mid \bar{q} q \mid 0 \rangle,\label{23}\ee
\be -i g \varepsilon_{\mu\nu\rho\tau} \langle 0 \mid \bar{q} \gamma_5 \frac{1}{2}
\lambda^n G^n_{\rho\tau} q \mid 0 \rangle_F =e_q \xi F_{\mu\nu} \langle 0 \mid \bar{q}q
\mid 0 \rangle\label{24}\ee Perform first the calculation of $\kappa$. In this case the
expression of $\psi_1(x,\rho)$ (\ref{16}) must be  multiplyed  by the additional factor:
$ \frac{1}{2} \tau^b G^b_{\mu\nu}$  where $G^b_{\mu\nu}$ is given by (\ref{11}) and the
indices $\mu,\nu$ in (\ref{16}) are changed to $\lambda,\sigma$. In the further
calculation it will  be taken into account, that $\chi_0$ in (\ref{7}) corresponds to
total spin-colour isospin $J=0$ and consequently
\be \vesig \vetau \chi_0 = -3\chi_0,~~~\sigma^a\tau^b \chi_0 = -\delta^{ab}
\chi_0.\label{26}\ee In the relation
\be \eta_{b\mu\nu} \eta_{b\lambda\sigma} = \delta_{\mu\lambda} \delta_{\nu\sigma}
-\delta_{\mu\sigma}\delta_{\nu\lambda} + \varepsilon_{\mu\nu\lambda\sigma}\label{27}\ee
the last term drops out after summation of zero-modes from instanton and anti-instanton
configuration. The final result for $\kappa$  is:
\be \kappa =-\int\limits^z_0 u^2 du \frac{1}{(u+1)^4} \biggl ( 1+\frac{1}{2}u\biggr )
=-\frac{1}{2} \biggl [ \ln (z+1) -\frac{13}{6} +\frac{1}{z+1} +\frac{1}{2}
\frac{1}{(z+1)^2} - \frac{1}{3} \frac{1}{(z+1)^3}\biggr ],\label{28}\ee where
$z=R^2/\rho^2 =12.7$. Numerically, we have:
\be \kappa = -0.26\label{29}\ee The calculation of $\xi$ is very similar to those of
$\kappa$  and the result is
\be \xi =2 \kappa =-0.52\label{30}\ee The values of $\kappa$ and $\xi$ only
logarithmically depend  on the cut-off. But unfortunately the logarithm in (\ref{28}) is
not very large and its main part is compensated by the term $-13/6$, appearing   in
(\ref{28}). So, the accuracy of (\ref{29}),(\ref{30}) can be estimated as about 30\%. The
phenomenological determination of 5-dimensional quark condensate magnetic
susceptibilities was performed by Kogan and Wyler \cite{Wyler} along the same lines, as
it was done in \cite{Kogan},\cite{Balit}. No anomalous dimensions  were accounted. The
results of \cite{Wyler} are:
\be \kappa =-0.34 \pm 0.1,~~~\xi =-0.74 \pm 0.2\label{31}\ee As can be seen, they are in
a good agreement with (\ref{29}),(\ref{30}). The 5-dimensional quark condensate magnetic
susceptibilities play a remarkable role in determination  of $\Lambda$-hyperon magnetic
moment \cite{Singh}.

I conclude this part of my talk. It was argued, that chiral symmetry violation in QCD
arises due to zero-mode solution of Dirac equation for massless quark in arbitrary gluon
field. The model is proposed similar to delute  instanton gas model, in which the
zero-mode solution is the same as in the field of instanton. The parameters of the model:
the density of zero-modes and their size are determined from the values of quark and
quark-gluon condensates. In the framework of this model  the values of quark condensates
magnetic susceptibilities of dimensions 3 and 5 were calculated in agreement with ones
found by QCD sum rules method using the properties of hadronic spectrum. The agreement of
these two approaches gives the strong argument in favour that the instantons are the only
source of chirality violating condensates in QCD.

Turn now to the temperature dependence of quark condensate, considered as an order
parameter. It is expected that quark condensate vanishes at high temperatures and the
chiral symmetry is restored. Two possibilities  are discussed: the second  order phase
transition  and the crossover. In our approach the normalization condition (\ref{8})
takes  place at any temperature \cite{Yaffe}. So, the temperature dependence of quark
condensate reduces to
\be \alpha (T) \equiv \langle 0 \mid \bar{q}q \mid \rangle_T = n (T)\label{36}\ee One may
expect that $n(T)$ vanishes only in the  case, when  the quantum number,  curried by
instanton is vanishing. This quantum number is  the topological charge, which is
temperature independent. So, in our approach $n(T)$ never vanishes and the  phase
transition  is of the type of crossover. The order parameter temperature dependence of
the second order phase transition near the critical point $T_c$ is smeared by
fluctuations \cite{Dau} (The same statement refers, surely, to the crossover.) The
fluctuation are determined by long wave oscillations of the fields. In QCD  they are
given by small frequency gluonic field  $G^n_{\mu\nu}$,  which is a constant. We have:
\be \frac{\Delta \alpha(T_c)}{\alpha(T_c)} \sim \langle 0 \mid G^2_{\mu\nu}\mid 0 \rangle
T^2_c \rho^6,\label{37}\ee where $\Delta \alpha(T_c)$ is of the order of magnitude of the
variation of $\alpha(T)$ near the crossover critical point. It is evident, that the left
hand side should vanish, if $T_c$ would be zero. I guess, that the factor, corresponding
to this circumstance is $T^2_c$. The factor $\rho^6$ is added for dimensional reasons.

All the  discussion of the $\alpha(T)$  temperature dependence is based on implicit
assumption, that this object has a physical meaning at finite $T$. It is unclear,
however, how it can be defined. I did not succeed to formulate the gedanken-experiments,
in which this object could  be measured at finite $T$. Therefore, may be, such an object
has no physical sence at $T\not =0$.

This work is supported by RFBR grant 09-02-00732 and in part by CRDF Cooperative Program
grant RUP2-2961-MO-09. I acknowledge the support of the European Community-Research
Infrastructure Integrating Activity ``Study of Strongly Interacting Matter'' under the
Seventh Framework Program of EU.

\end{document}